\documentclass[aps,prb,twocolumn,showpacs,superscriptaddress,floatfix]{revtex4-1}
\usepackage{amsmath,amssymb}
\usepackage{graphicx}
\usepackage{color}
\usepackage{hyperref}
\usepackage{bm}
\usepackage{setspace}
\usepackage{appendix}

\begin{document}

\title{Dopant clustering, electronic inhomogeneity, and vortex pinning in iron-based superconductors}

% Electronic Inhomogeneity and Vortex Pinning in Fe-based Superconductors
% Dopant Contributions to Vortex Pinning in Fe-based Superconductors
% Dopant Clustering and Vortex Disorder in Fe-based Superconductors
% Dopant Clustering, Electronic Inhomogeneity, and Vortex Pinning in Fe-based Superconductors

\author{Can-Li Song}
\affiliation{Department of Physics, Harvard University, Cambridge, MA 02138, USA}
\author{Yi Yin}
\affiliation{Department of Physics, Harvard University, Cambridge, MA 02138, USA}
\affiliation{Physics Department, Zhejiang University, Hangzhou, 310027, China}
\author{Martin Zech}
\author{Tess Williams}
\author{Michael M. Yee}
\affiliation{Department of Physics, Harvard University, Cambridge, MA 02138, USA}
\author{Gen-Fu Chen}
\author{Jian-Lin Luo}
\author{Nan-Lin Wang}
\affiliation{Beijing National Laboratory for Condensed Matter Physics and Institute of Physics, Chinese Academy of Sciences, Beijing 100190, China}
\author{E. W. Hudson}
\affiliation{Department of Physics, The Pennsylvania State University, University Park, PA 16802, USA}
\author{Jennifer E. Hoffman}
\email[]{jhoffman@physics.harvard.edu}
\affiliation{Department of Physics, Harvard University, Cambridge, MA 02138, USA}
\date{\today}

\begin{abstract}
We use scanning tunneling microscopy to map the surface structure, nanoscale electronic inhomogeneity, and vitreous vortex phase in the hole-doped superconductor Sr$_{0.75}$K$_{0.25}$Fe$_2$As$_2$ with $T_c$=32 K. We find the low-$T$ cleaved surface is dominated by a half-Sr/K termination with $1\times 2$ ordering and ubiquitous superconducting gap, while patches of gapless, unreconstructed As termination appear rarely. The superconducting gap varies by $\sigma/\overline{\Delta}$=16\% on a $\sim$3 nm length scale, with average $2\overline{\Delta}/k_B T_c=3.6$ in the weak coupling limit. The vortex core size provides a measure of the superconducting coherence length $\xi$=2.3 nm. We quantify the vortex lattice correlation length at 9 T in comparison to several iron-based superconductors. The comparison leads us to suggest the importance of dopant size mismatch as a cause of dopant clustering, electronic inhomogeneity, and strong vortex pinning.
\end{abstract}

\maketitle
% body of paper here
\section{\label{sec:Intro}Introduction}
The recent discovery of high transition temperature ($T_{c}$) iron-based superconductors (Fe-SCs)\cite{kamihara2008iron} has provoked tremendous excitement in condensed matter physics, and launched a new era in the search for the key to high-$T_{c}$ superconductivity.\cite{mazin2010superconductivity} Like cuprates, Fe-SCs exhibit a layered structure with electronically active superconducting planes separated by buffer layers, and the superconductivity develops from antiferromagnetic parent compounds upon chemical doping. In addition to enabling superconductivity, the dopants are potential sources of nanoscale phase separation,\cite{park2009electronic} and crystalline\cite{massee2009cleavage} and electronic disorder,\cite{yin2009scanning,julien2009homogeneous,massee2009nanoscale} which may in turn lead to $T_{c}$ suppression\cite{tropeano2010isoelectronic} and vortex pinning.\cite{van2012vortex} With the diversity of possible dopants in Fe-SCs, it has remained an elusive challenge to characterize and categorize their nanoscale effects.

With its atomic scale structural and spectroscopic imaging abilities, the scanning tunneling microscope (STM) has proven to be an ideal tool to study the nanoscale properties of correlated electron materials. However, STM studies of Fe-SCs have presented several controversial results.\cite{hoffman2011spectroscopic} First, the cleaved $A$Fe$_{2}$As$_{2}(001)$ surface showed both $1\times 2$ and $\sqrt{2}\times\sqrt{2}$ reconstructions with unclear origin: either a half-layer of \textit{A} or a reconstruction of the complete As layer. Second, spectroscopic images of optimally electron-doped Ba(Fe$_{1-x}$Co$_{x}$)$_{2}$As$_{2}$ cleaved at $\sim$25 K revealed nanoscale variations in the superconducting gap $\Delta$ on a length scale of several nanometers.\cite{yin2009scanning} However, studies of the same compound cleaved at room temperature found a shorter $\Delta$ correlation length of $\sim$1.0 nm, closely matching the average Co separation for a random dopant distribution, which prompted the hypothesis that the gap variations were caused by the disorder of individual Co atoms.\cite{massee2009nanoscale} Third, electron-doped $\textrm{Ba}\textrm{Fe}_{1.8}\textrm{Co}_{0.2}\textrm{As}_{2}$ displayed a disordered vortex lattice without observed Andreev bound states at vortex core.\cite{yin2009scanning} In contrast, hole-doped Ba$_{0.6}$K$_{0.4}$Fe$_{2}$As$_{2}$ displayed a hexagonal vortex lattice with pronounced vortex core bound states.\cite{shan2011observation} It has been proposed but not verified that the vortex discrepancy may be explained by stronger scattering from the in-plane Co dopants than the out-of-plane K dopants.

To address these controversies, the hole-doped Sr$_{1-x}$K$_{x}$Fe$_{2}$As$_{2}$ [Fig.\ \ref{fig:structure}(a)] is a unique system with specific advantages.\cite{chen2008transport} First, Gao \textit{et al} predicted that in contrast to BaFe$_{2}$As$_{2}$, As-terminated SrFe$_{2}$As$_{2}$ would show no surface reconstruction.\cite{gao2010surface} If this prediction holds true, it should allow easy distinction between a complete As layer and a partial Sr layer at the cleaved surface of Sr$_{1-x}$K$_{x}$Fe$_{2}$As$_{2}$. More importantly, Sr$_{1-x}$K$_{x}$Fe$_{2}$As$_{2}$ serves as a test case -- a tie-breaker of sorts -- to understand the gap inhomogeneity and vortex pinning differences between BaFe$_{1.8}$Co$_{0.2}$As$_{2}$ and Ba$_{0.6}$K$_{0.4}$Fe$_{2}$As$_{2}$.\cite{yin2009scanning, shan2011observation}  Here we use STM to investigate slightly underdoped Sr$_{0.75}$K$_{0.25}$Fe$_2$As$_2$ single crystals with $T_{c}$=32 K.  We emphasize that our Sr$_{0.75}$K$_{0.25}$Fe$_2$As$_2$ crystals are grown with similar flux methods as the BaFe$_{1.8}$Co$_{0.2}$As$_{2}$ and Ba$_{0.6}$K$_{0.4}$Fe$_{2}$As$_{2}$ crystals studied earlier,\cite{yin2009scanning,Wang2009f,Choi2009,Wang2009g,shan2011observation, } facilitating direct comparison (see Table 1 in the Supplemental Material).\cite{supplementary}

\section{\label{sec:Exper}Experimental}
High quality Sr$_{0.75}$K$_{0.25}$Fe$_{2}$As$_{2}$ single crystals were grown by the FeAs flux method.\cite{chen2008transport} FeAs was obtained by reacting a mixture of powered elements in an evacuated quartz tube. Mixtures of Sr, K, and FeAs powders were then put into an alumina crucible and sealed in a welded Ta crucible with Ar gas. The Ta crucible is sealed in an evacuated quartz ampoule and heated at $1150^{\circ}$C for 5 hours and cooled slowly to 800$^{\circ}$. Platelike crystals with size up to a centimeter could be obtained according to this recipe.\cite{chen2008transport}

All experiments are carried out using a home-built cryogenic STM. Samples are cleaved \textit{in situ} at $\sim$25 K and inserted immediately into the STM for imaging at 6 K. Mechanically cut polycrystalline PtIr tips are sharpened by field emission, and screened for featureless density of states on an Au target. To obtain a tunneling current, a bias is applied to the sample while the tip is held at virtual ground. Tunneling conductance (which is proportional to the local density of states) is measured using a standard lock-in technique with a 1.0 mV rms bias modulation at 1110 Hz. The magnetic field up to 9 Tesla is applied perpendicular to the sample surface.

\section{\label{sec:Res}Results}
\subsection{\label{sec:surf}Surface structure}
\begin{figure}[tb]
\center
\includegraphics[width=0.89\columnwidth]{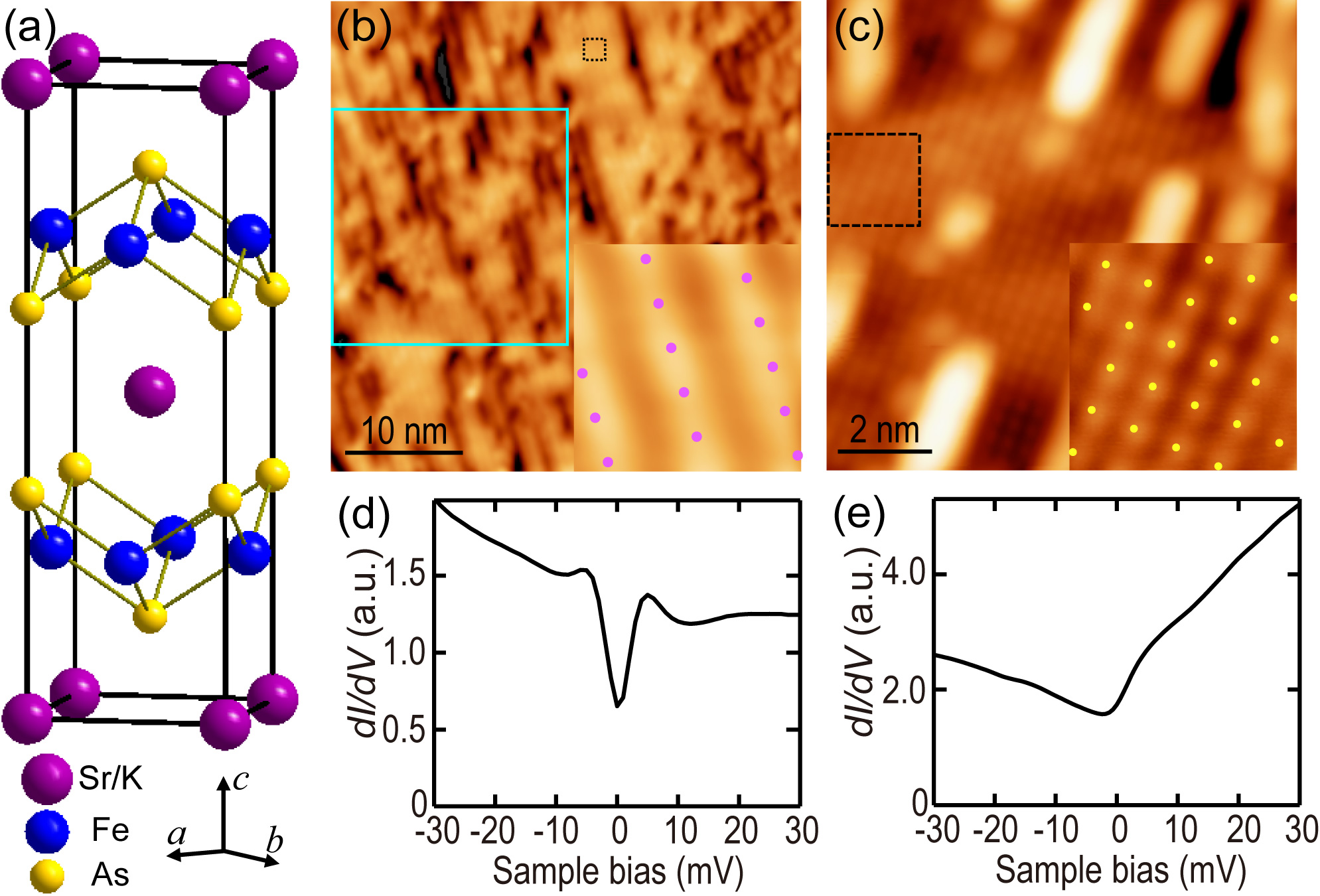}
\caption{(color online) (a) Schematic crystal structure of Sr$_{1-x}$K$_{x}$Fe$_{2}$As$_{2}$. (b) STM topography ($V_{\mathrm{s}}$ = -100 mV, $I$ = 35 pA) of the commonly observed Sr$_{0.75}$K$_{0.25}$Fe$_{2}$As$_{2}$ surface with 1 $\times$ 2 stripe order. Inset shows a zoom-in of the stripe (black square), with the magenta dots denoting Sr/K atoms ($V_{\mathrm{s}}$ = -100 mV, $I$ = 30 pA, 2 nm $\times$ 2 nm). The cyan box indicates the region where $dI/dV$ spectra were acquired for the maps in Figs.\ 2(a)-(c). (c) STM topography ($V_{\mathrm{s}}$ = -100 mV, $I$ = 350 pA) showing As-terminated $1 \times 1$ surface, decorated by sparse Sr/K rows. Inset shows a magnification of the 1 $\times$ 1 surface (black square), with the yellow dots denoting As atoms ($V_{\mathrm{s}}$ = -100 mV, $I$ = 350 pA, 2 nm $\times$ 2 nm). (d, e) Spatially averaged \textit{dI/dV} spectra in (b) and (c) regions, respectively. Tunneling gap was stabilized at $V_{\mathrm{s}}$ = -100 mV and $I$ = 300 pA.}
\label{fig:structure}
\end{figure}

Figure \ref{fig:structure}(b) shows a topographic image of a typical $\sim$25 K-cleaved Sr$_{0.75}$K$_{0.25}$Fe$_2$As$_2$ surface, displaying local $1\times 2$ stripes.\cite{niestemski2009unveiling, boyer2008scanning} Due to the stronger bonding within the FeAs layers, the cleavage likely occurs within the Sr/K plane, leaving approximately half of the Sr/K atoms on either exposed side to balance the chemical valence. Occasionally for low-$T$ cleaves, the metastable As-terminated $1 \times 1$ surface may be expected.\cite{gao2010surface} However, previous STM images of SrFe$_{2}$As$_{2}$ samples cleaved at 77 K showed no such $1 \times 1$ patches,\cite{niestemski2009unveiling} and instead both $1\times 2$ and $\sqrt{2}\times\sqrt{2}$ orders covering the entire cleaved surface were explained as the bare but reconstructed As layer. Another earlier study of Sr$_{1-x}$K$_{x}$Fe$_{2}$As$_{2}$ cleaved at 10 K showed only a very small 1 $\times$ 1 patch with no accompanying spectroscopy.\cite{boyer2008scanning} In contrast, we present the first observation of larger 1 $\times$ 1 patches [Fig.\ \ref{fig:structure}(c)], constituting $\sim$5$\%$ of the surface, with average As-As atomic spacing of approximately 0.4 nm. This observation provides evidence that the dominant $1\times 2$ and $\sqrt{2}\times\sqrt{2}$ structures arise from a half-Sr/K layer. The bright rows in our 1 $\times$ 1 regions are identified as residual Sr/K atoms.\cite{Dreyer2011}

We record differential $dI/dV$ spectra on both Sr/K- and As-terminated surfaces, illustrated in Figs.\ \ref{fig:structure}(d) and \ref{fig:structure}(e), respectively. In stark contrast to the universal superconducting gap with clear coherence peaks on the Sr/K-terminated surface, no superconducting gap is observed on the As-terminated surface. This is probably caused by the strong polarity of the latter surface, which causes the surface to deviate from the doping which supports the superconductivity. The cleaved structure therefore plays a crucial role in superconductivity at the surface, in some cases preventing even proximity-induced superconductivity from appearing due to the short $c$-axis coherence length $\xi$.

\subsection{\label{sec:inhomogeneity}Electronic inhomogeneity}

\begin{figure}[tb]
\center
 \includegraphics[width=\columnwidth]{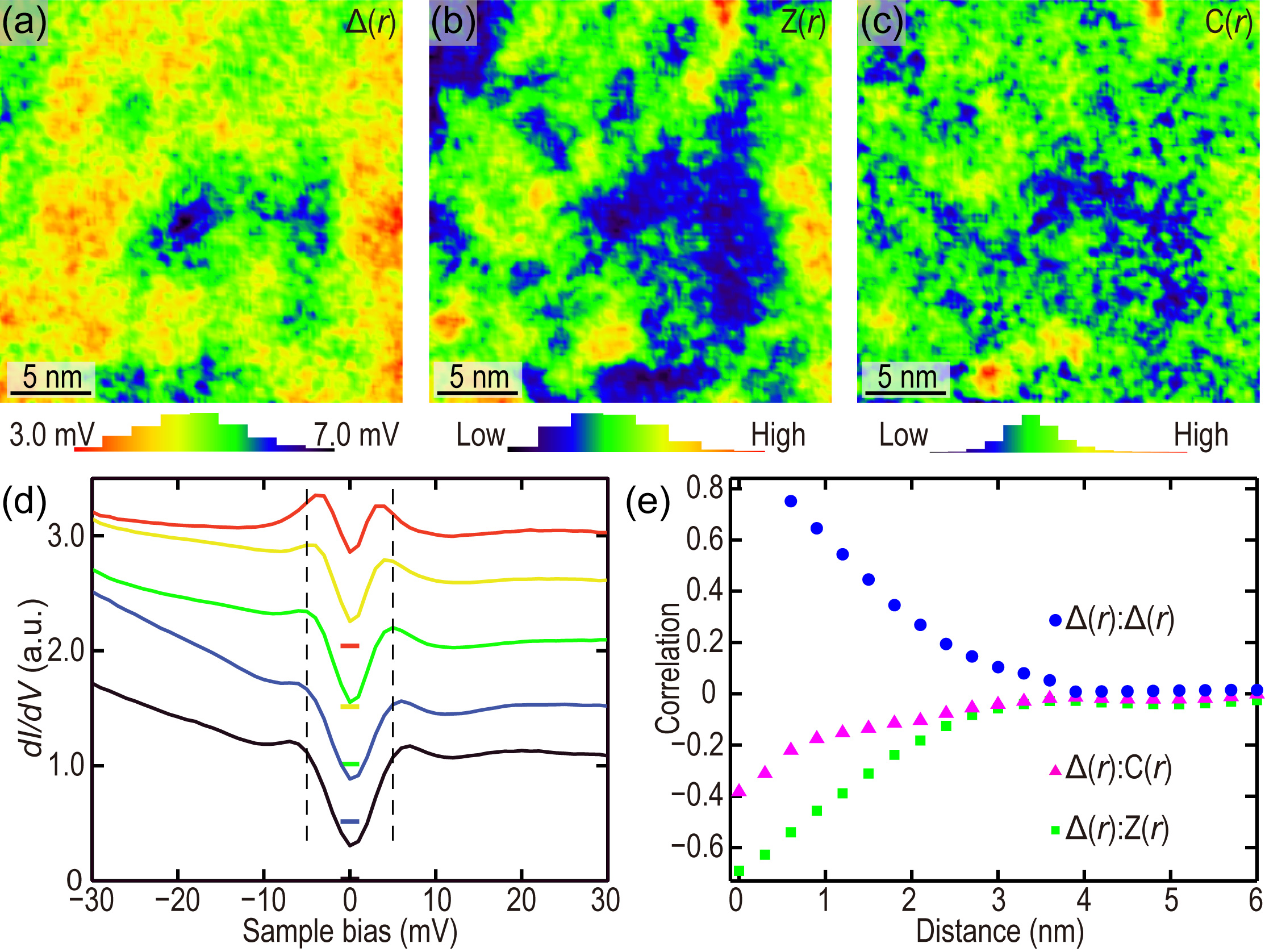}
  \caption{(color online) (a-c) Maps of $\Delta(\vec{r})$ (half the distance between coherence peaks), $Z(\vec{r})$, and $C(\vec{r})$ (average conductance at the two coherence peaks). (d) Binned and averaged raw spectra for five ranges of gap $\Delta$ ($\Delta_{\mathrm{min}}$-3.8, 3.8-4.7, 4.7-5.5, 5.5-6.3, 6.3-$\Delta_{\mathrm{max}}$ meV from top to bottom), color-coded to match those in (a). The spectra have been vertically offset for clarity, with their zero-conductance positions marked by correspondingly colored horizontal lines. Vertical dashes at $\pm$5 meV are guides to the eye. (e) Azimuthally averaged autocorrelation of $\Delta(\vec{r})$, and cross-correlations of $\Delta(\vec{r})$ with $Z(\vec{r})$ and $C(\vec{r})$.}
\label{fig:gapmap}
\end{figure}

To further explore the superconducting surface, we survey $dI/dV$ spectra within the cyan box in Fig.\ \ref{fig:structure}(b), which can be analyzed to yield maps of $\Delta(\vec{r})$, zero bias conductance (ZBC) $Z(\vec{r})$, and coherence peak strength $C(\vec{r})$, as summarized in Figs.\ \ref{fig:gapmap}(a)-(c). Here all spectra have been normalized by their backgrounds to compute $Z(\vec{r})$ and $C(\vec{r})$, as detailed in Figs.\ S1 and S2.\cite{supplementary} All three maps show spatial inhomogeneity. The average $\overline{\Delta}$ = 5.0 meV with a standard deviation of $\sigma$ = 0.8 meV gives a fractional variation $\sigma/\overline{\Delta}$ = $16\%$, larger than the $12\%$ variation found in BaFe$_{1.8}$Co$_{0.2}$As$_{2}$.\cite{yin2009scanning} The small reduced gap $2\overline{\Delta}/k_B T_c \sim$3.6 suggests weaker coupling in Sr$_{0.75}$K$_{0.25}$Fe$_{2}$As$_{2}$.

To investigate the spectral shape, all spectra are binned based on $\Delta$, with their averages plotted in Fig.\ \ref{fig:gapmap}(d). Evidently, the spectra with larger $\Delta$ tend to show smaller ZBC and weaker coherence peaks, similar to the gap-coherence relation observed in cuprates,\cite{lang2002imaging} which may stem from a scattering rate that increases with energy.\cite{alldredge2008evolution} Such broadening possibly originates from the competition of superconductivity with the inhomogeneous spin density wave gap in Fe-SCs.\cite{niestemski2009unveiling, gen2008superconductivity, cai2013visualizing} To quantify this trend, we compute the cross-correlations between $\Delta(\vec{r})$, $Z(\vec{r})$, and $C(\vec{r})$ [Fig.\ \ref{fig:gapmap}(e)]. Together with the $\Delta$ autocorrelation, we note that the characteristic length scale over which the correlations go to zero is $\sim$3 nm, exceeding the average separation ($\sim$1.1 nm) between individual K dopants. We therefore hypothesize that the K dopants exhibit nanoscale phase separation, clustering to form K-rich and K-poor regions,\cite{yeoh2011direct} although further STM conductance mapping experiments at higher energies would likely be needed in order to directly image these K clusters.\cite{nanoscale2013Zeljkovic} The formation of clusters can more effectively relax the strain caused by the large ion size mismatch between K$^{+}$ (146 pm) and Sr$^{2+}$ (126 pm).\cite{shannon197revised} Due to the high sensitivity of superconductivity to the chemical doping,\cite{gen2008superconductivity} the resultant K clusters could account well for the observed electronic inhomogeneity.

\subsection{\label{sec:vlattice}Vortex arrangement}
\begin{figure}[tb]
\center
  \includegraphics[width=0.97\columnwidth]{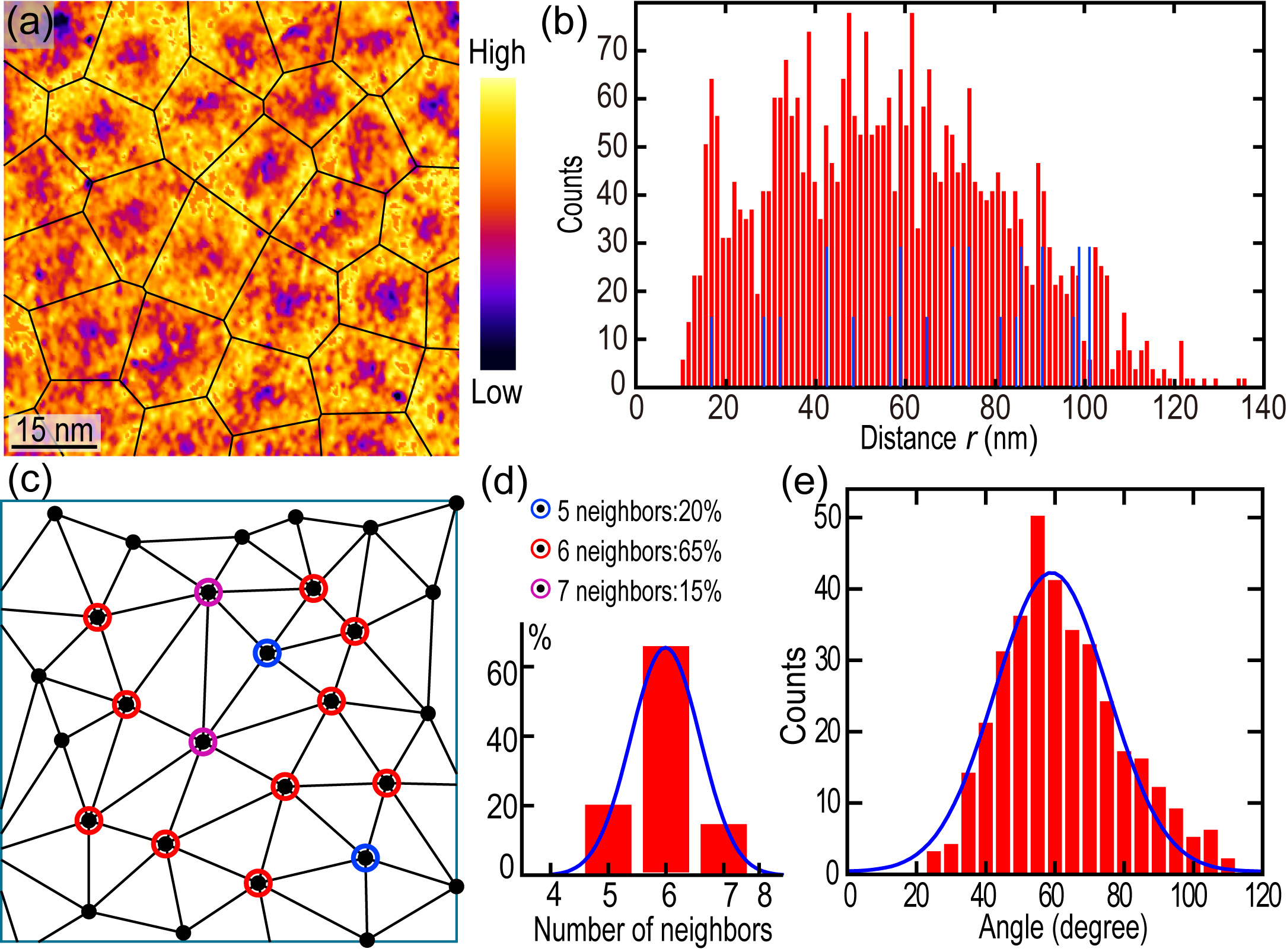}
  \caption{(color online) (a) A 80 nm $\times$ 80 nm $dI/dV$ map at $V_{\mathrm{s}}$ = -5 meV, recorded in a 9 T $c$-axis magnetic field. The vortex centers are determined by 2D Gaussian fitting. (b) Histograms of the vortex pair distances $d_{ij}$ from (a) and another similar 100 nm $\times$ 100 nm vortex map. Vertical blue lines correspond to the positions and relative weight of pair distances for an ideal triangular vortex lattice at 9 T. (c) Delaunay triangulation (black lines) of the vortex lattice in (a). Each vortex is color-coded based on the number of its nearest neighbors. (d, e) Coordination number and Delaunay angle distributions on the border-free regions. Solid blue lines are the Gaussian fits.}
\label{fig:Vlattice}
\end{figure}

Magnetic vortices are technologically important, as the superconducting critical current $J_{c}$ is limited by the vortex pinning strength. The characterization of vortices is also scientifically valuable for determining the superconducting coherence length $\xi$\cite{yin2009scanning, shan2011observation} and pairing symmetry.\cite{song2011direct} We image the vortices in Sr$_{0.75}$K$_{0.25}$Fe$_{2}$As$_{2}$ by mapping $dI/dV$ at the filled state coherence peak (-5 meV), shown in Fig.\ \ref{fig:Vlattice}(a). The vortices locally suppress the superconducting coherence peaks, and appear as purple-black features with depressed conductance. To better emphasize the vortices, Voronoi cells are overlaid onto the image.\cite{yin2009scanning} From the Voronoi cell size, we estimate the average flux per vortex $\Phi = 2.1\pm0.1\times 10^{-15}$ Wb, consistent with one magnetic flux quantum, $\Phi_{0} = 2.07 \times 10^{-15}$ Wb.

In stark contrast to the hexagonal vortex lattice which indicates negligible pinning in Ba$_{0.6}$K$_{0.4}$Fe$_{2}$As$_{2}$,\cite{shan2011observation} the vortices in Sr$_{0.75}$K$_{0.25}$Fe$_{2}$As$_{2}$ do not form an ordered lattice. Instead, similar to electron-doped Ba(Fe$_{1-x}$Co$_x$)$_2$As$_2$,\cite{yin2009scanning, inosov2010symmetry} a short-range hexagonal order (vortex glass phase) is justified based on the following two tests. First, Fig.\ \ref{fig:Vlattice}(b) shows a histogram of relative distances $d_{ij} = |r_{i} - r_{j}|$, calculated for all vortex pairs at positions $r_{i}$ and $r_{j}$. The pronounced peak at the smallest distance of $\sim$16.4 nm coincides almost exactly with the expected lattice constant $a=\sqrt{2\Phi_{0}/\sqrt{3}H}$ ($\sim$16.3 nm, the first vertical blue line) for a perfect hexagonal vortex arrangement at 9 T. However, with increasing distance the experimental histogram peaks cease to match the corresponding hexagonal lattice separations. This presents the first indication that the vortices are of short-range hexagonal order. Second, Fig.\ \ref{fig:Vlattice}(c) shows the Delaunay triangulation by connecting all nearest neighbor vortex sites.\cite{iavarone2008effect, inosov2010symmetry, hanaguri2012scanning} Each vortex, surrounded by a closed Voronoi polygon,\cite{inosov2010symmetry} is color-coded by its coordination number. The statistics of the coordination number and angles of Delaunay triangles are plotted in Figs.\ \ref{fig:Vlattice}(d) and \ref{fig:Vlattice}(e), respectively. Two-thirds of vortices are six-fold coordinated, and more importantly, the Delaunay angle distribution shows a single pronounced peak at $\sim60^{\circ}$. All the evidence consistently supports a short-range hexagonal order of the vortex arrangement, indicative of strong vortex pinning in Sr$_{0.75}$K$_{0.25}$Fe$_{2}$As$_{2}$.

\subsection{\label{sec:vcore}Vortex core}

\begin{figure}[tb]
\center
 \includegraphics[width=0.73\columnwidth]{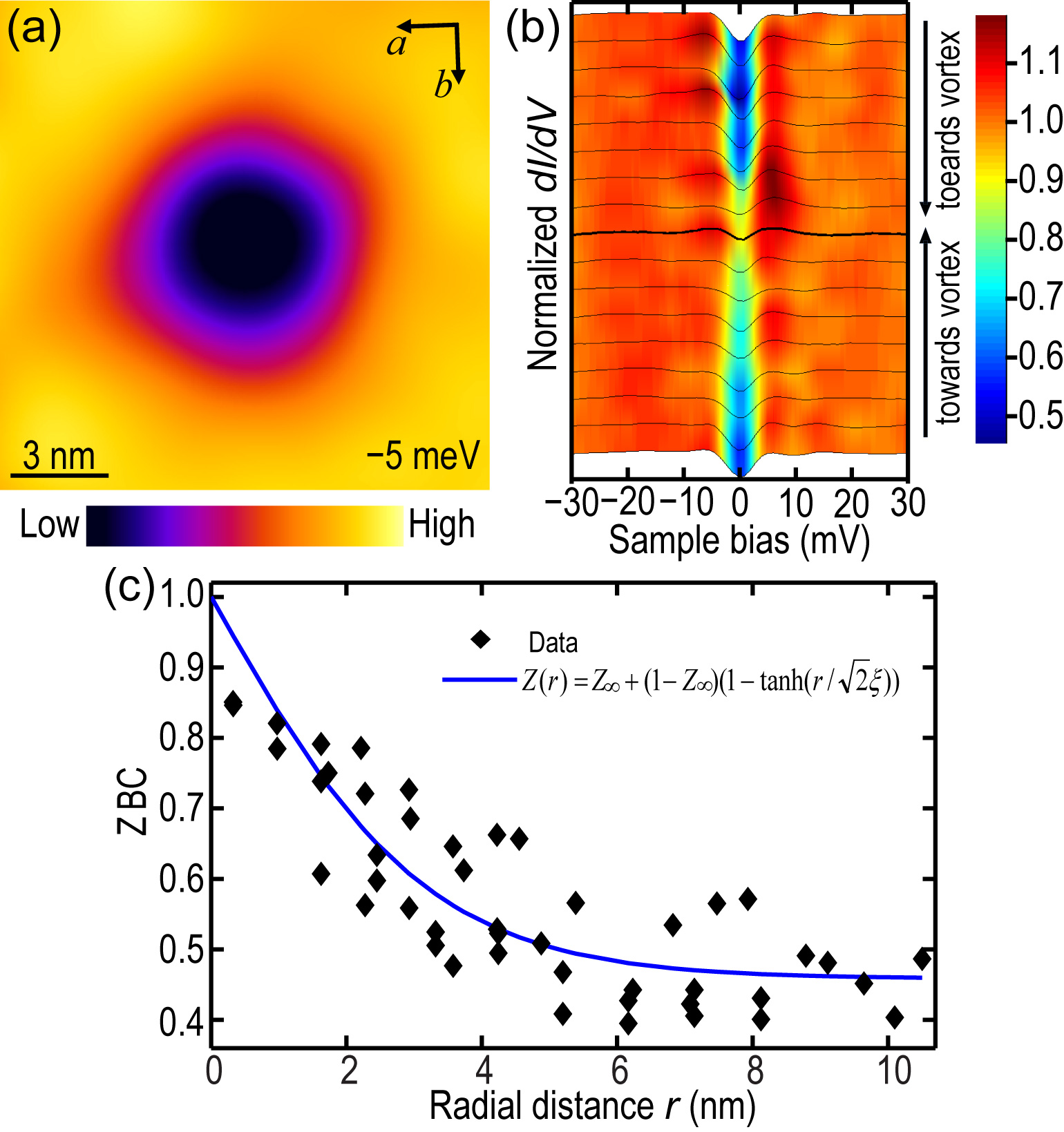}
  \caption{(color online) (a) Average $dI/dV$ at -5 mV from overlaying 48 single vortices, revealing an isotropic vortex core. (b) Typical series of normalized $dI/dV$ spectra straddling a single vortex, with the thicker one near the vortex core. The spectra are equally separated and span a total distance of 11 nm. (c) Radial dependence of $Z(\vec{r})$ around three vortices. Blue line shows the best fit of $Z(\vec{r})$ to equation (\ref{eq:GLvortex}).}
\label{fig:Vcore}
\end{figure}

To investigate the shape of the vortex core, we register all vortex centers, then average the density of states around 48 vortices, as depicted in Fig.\ \ref{fig:Vcore}(a). Such averaging should enhance any intrinsic vortex core shape due to band structure or pairing anisotropy,\cite{song2011direct, hanaguri2012scanning, Wang2012theory} while minimizing extrinsic effects from the pinning-induced variations in
coordination number and nearest-neighbor directions. In contrast to the two- and four-fold symmetric vortices in FeSe\cite{song2011direct} and LiFeAs,\cite{hanaguri2012scanning} the vortices in Sr$_{0.75}$K$_{0.25}$Fe$_{2}$As$_{2}$ are nearly isotropic. This observation does not support the claims of \textit{d}-wave pairing in more overdoped $A_{1-x}$K$_{x}$Fe$_{2}$As$_{2}$ materials,\cite{reid2012universal} although it leaves open the possibility that the isotropy stems from thermal smearing\cite{hanaguri2012scanning} or impurity scattering.\cite{pan2000stm}

A series of normalized \textit{dI/dV} spectra across one vortex are shown in Fig.\ \ref{fig:Vcore}(b). As expected, $Z(\vec{r})$ is elevated within the vortex core, reflecting the suppression of superconductivity there. The vortex-induced $Z(\vec{r})$ as a function of the radial distance from the vortex center is shown in Fig.\ \ref{fig:Vcore}(c). From the Ginzburg-Landau expression for the superconducting order parameter $\Psi(r)$ near the interface between a superconductor and a normal metal, the ZBC profile across vortex core should obey
\begin{equation} \label{eq:GLvortex}
Z(r)=Z_{\infty}+(1-Z_{\infty})(1-\textrm{tanh}(-r/\sqrt{2}\xi))\,,
\end{equation}
where $Z_{\infty}$ is the normalized ZBC away from the vortex core and $r$ the distance to the vortex center (see the details in Fig.\ S3).\cite{Eskildsen2002vortex, Bergeal2006scanning, ning2010vortex, zhang2010superconductivity, supplementary} A fit to equation (\ref{eq:GLvortex}) yields a superconducting coherence length $\xi=2.3\pm0.2$ nm, which matches excellently $\xi_{ab}=2.1$ nm in Sr$_{0.6}$K$_{0.4}$Fe$_{2}$As$_{2}$ from transport measurements.\cite{marra2012paraconductivity}

\section{\label{sec:Dis}Discussion}
\begin{figure}[b]
\center
 \includegraphics[width=0.91\columnwidth]{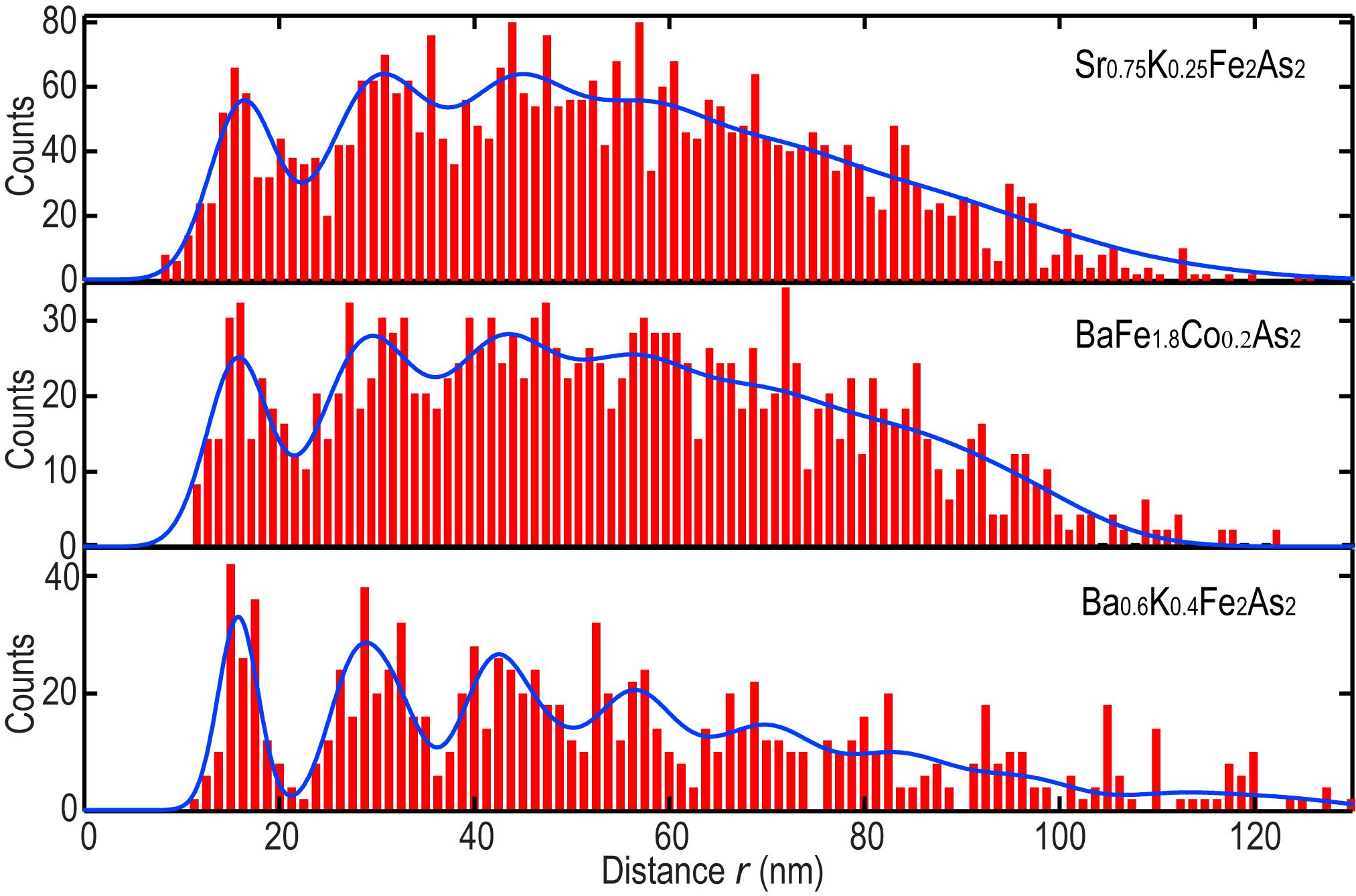}
  \caption{(color online) Histograms of the vortex pair distances $d_{ij}$ in three Fe-SCs. Values extracted from vortex lattices in BaFe$_{1.8}$Co$_{0.2}$As$_{2}$\cite{yin2009scanning} and Ba$_{0.6}$K$_{0.4}$Fe$_{2}$As$_{2}$\cite{shan2011observation} are added for comparison. Blue line shows the best RDF fits to equation (\ref{eq:edgeeffects})\cite{inosov2010symmetry}.}
\label{fig:RDF}
\end{figure}

\begingroup
\begin{table*}[tbh]
\center
\caption{\label{tab:ionsize}Best RDF fitting parameters of vortex distance $d_{ij}$, ion size and vortex arrangement for several Fe-SCs. The lattice constant $a_{\vartriangle}$=$\sqrt{2\Phi_{0}/\sqrt{3}H}$, expected for a perfect triangular vortex lattice at 9 T, are shown in the second column. Note that the measured radius of the first coordination shell $R_{1}$ is compared with $a_{\vartriangle}$. All values are given in nm, unless otherwise specified. The statistical errors of $\zeta$ indicates the standard derivation of $\zeta$ values obtained for different binning of the histograms. The value $\zeta$ = 22 nm in BaFe$_{1.8}$Co$_{0.2}$As$_{2}$ agrees with that obtained by Inosov \textit{et al}\cite{inosov2010symmetry}. }
%\begin{footnotesize}
\scalebox{1}{
\begin{tabular}{ l c c c c c c c c c c r }
  \hline
    & & & & & & & Native atom & Dopant & & & \\
  Fe-SC  & $N$ & $a_{\vartriangle}$ & $R_1$ & $\sigma$ & $R_{\mathrm{max}}$ & $W_{\mathrm{max}}$ & size (pm) & size (pm) & Diff. (pm) & $\zeta$ & Ref. \\
  \hline\hline
  Sr$_{0.75}$K$_{0.25}$Fe$_{2}$As$_{2}$  &82& 16.3 & 16.5 & 3.42 &  46 & 21 & 126 & 146 & 20 & $20\pm1$ &  \\
  BaFe$_{1.8}$Co$_{0.2}$As$_{2}$  &41 &  16.3 & 15.9 & 3.25 &39 & 25 & 78  &  61 & 17 & $22\pm1$ & \onlinecite{yin2009scanning} \\
  Ba$_{0.6}$K$_{0.4}$Fe$_{2}$As$_{2}$    &34& 16.3 &  16.0 & 2.08 & 31 & 23 & 142 & 146 &  4 & $52\pm2$ & \onlinecite{shan2011observation} \\
  FeSe    &--&--&--&--&--&--& 198 & n/a &  0 & ordered & \onlinecite{song2011direct} \\
  FeSe$_{0.4}$Te$_{0.6}$ &--&--&--&--&--&--& 198 & 221 &  23 & disordered & \onlinecite{hanaguri2010unconventional} \\
  \hline
\end{tabular}}
%\end{footnotesize}
\end{table*}
\endgroup

By directly comparing our Sr$_{0.75}$K$_{0.25}$Fe$_{2}$As$_{2}$ measurements to earlier work on  BaFe$_{1.8}$Co$_{0.2}$As$_{2}$ and Ba$_{0.6}$K$_{0.4}$Fe$_{2}$As$_{2}$, we arrive at some general suggestions about vortex pinning in Fe-SCs. We consider several hypotheses for the differences between the vortex arrangements in these three materials.
First, it has been argued that collective pinning of vortices in Fe-SCs arises from charge doping.\cite{van2010quasiparticle} However, the charges of the K$^+$ dopants and the Sr$^{2+}$ and Ba$^{2+}$ ions they replace are identical in Sr$_{1-x}$K$_{x}$Fe$_{2}$As$_{2}$ and Ba$_{1-x}$K$_{x}$Fe$_{2}$As$_{2}$, so the charge model alone cannot easily explain the contrast between our observed strong vortex pinning in Sr$_{0.75}$K$_{0.25}$Fe$_{2}$As$_{2}$ and the ordered vortices which indicate weak pinning in Ba$_{0.6}$K$_{0.4}$Fe$_{2}$As$_{2}$. Second, one may expect that the shear modulus $C_{66}$, roughly proportional to $H_{c2}^{2}b(1-b)^{2}$ (where $b=H/H_{c2}$),\cite{brandt1986elastic} will play a role in the different vortex arrangements between Ba$_{0.6}$K$_{0.4}$Fe$_{2}$As$_{2}$ and Sr$_{0.75}$K$_{0.25}$Fe$_{2}$As$_{2}$. A larger $C_{66}$ often results in an ordered vortex lattice. Using the Ginzburg-Landau expression $H_{c2}=\Phi_{0}/2\pi\xi^{2}$ and $\xi=2.3\pm0.2$ nm, we estimate $H_{c2}$ = 62 $\pm$ 11 T in Sr$_{0.75}$K$_{0.25}$Fe$_{2}$As$_{2}$, quite close to $H_{c2}\sim$ 75 T in Ba$_{0.6}$K$_{0.4}$Fe$_{2}$As$_{2}$.\cite{shan2011observation} Moreover, the persistence of vortex lattice order in Ba$_{0.6}$K$_{0.4}$Fe$_{2}$As$_{2}$ down to 4 T\cite{shan2011observation} (where $C_{66}$ is even smaller than that of Sr$_{0.75}$K$_{0.25}$Fe$_{2}$As$_{2}$ at 9 T) suggests that $C_{66}$ cannot solely account for the short-range vortex order in Sr$_{0.75}$K$_{0.25}$Fe$_{2}$As$_{2}$. Finally, we consider the nanoscale electronic inhomogeneity [Fig.\ \ref{fig:gapmap}], which we hypothesized to be caused by K clustering. Since the ion size mismatch between K$^{+}$ and Ba$^{2+}$ is five times smaller than that between K$^{+}$ and Sr$^{2+}$ (Table \ref{tab:ionsize}), one can expect K$^{+}$ ions to be less clustered in Ba$_{0.6}$K$_{0.4}$Fe$_{2}$As$_{2}$ than in Sr$_{0.75}$K$_{0.25}$Fe$_{2}$As$_{2}$.\cite{zajdel2010phase} This leads to smaller inhomogeneity and weaker vortex pinning, consistent with the well-ordered vortex lattices in Ba$_{0.6}$K$_{0.4}$Fe$_{2}$As$_{2}$.\cite{shan2011observation} In contrast, the large ion size mismatch between K$^{+}$ and Sr$^{2+}$ could account for the greater electronic inhomogeneity and disordered vortex lattice in Sr$_{0.75}$K$_{0.25}$Fe$_{2}$As$_{2}$ observed here.

To quantify the effect of the ion size mismatch, we analyze the radial distribution function (RDF) of the vortex lattices,  $f(r)$, in several similar Fe-SCs grown by the similar flux method.\cite{yin2009scanning,shan2011observation,inosov2010symmetry} The RDF can be well approximated by a sum of Gaussian peaks, with widths increasing proportionally to $\sqrt{r}$,
\begin{equation} \label{eq:sumgauss}
f(r)=\sum_{n=1}^{\infty} \frac{N_n}{\sigma \sqrt{2\pi R_n/a_{\vartriangle}}} \exp\left[ - \frac{(r-R_n)^2}{2\sigma^2 R_n/a_{\vartriangle}} \right]\,
\end{equation}
Here $a_{\vartriangle}$=$\sqrt{2\Phi_{0}/\sqrt{3}H}$ is the lattice constant expected for a perfect triangular vortex lattice at a magnetic field of \textit{H}, $\sigma \ll a_{\vartriangle}$ is the standard deviation of the difference between
nearest-neighbor distances and $a_{\vartriangle}$, $R_n$ is the radius of the $n^{\mathrm{th}}$ coordination shell, and $N_n$ is the number of sites in this shell. On short length scales, the RDF will exhibit oscillatory behavior, with roughly exponentially decaying amplitude $\sim \exp(-r/\zeta)$, where $\zeta$ is defined as the radial correlation length of the vortex lattice, and is given by\cite{inosov2010symmetry}
\begin{equation} \label{eq:zetadef}
\zeta = \frac{\sigma^2}{a_{\vartriangle}} \left[ \sqrt{\frac{1}{2} + \sqrt{\frac{1}{4} + 4\pi^2 \frac{\sigma^4}{a_{\vartriangle}^4}}} -1 \right]^{-1}\,
\end{equation}
Larger $\zeta$ always corresponds to a more ordered vortex lattice with weaker vortex pinning.

Histograms of the observed $d_{ij}$ in Sr$_{0.75}$K$_{0.25}$Fe$_{2}$As$_{2}$, Ba(Fe$_{1-x}$Co$_x$)$_2$As$_2$\cite{yin2009scanning} and Ba$_{0.6}$K$_{0.4}$Fe$_{2}$As$_{2}$\cite{shan2011observation} are plotted in Fig.\ \ref{fig:RDF}. The blue lines show the fits to
\begin{equation} \label{eq:edgeeffects}
\frac{ \ N f(r) \delta r} {1+ \exp \left[ (r-R_{\mathrm{max}})/W_{\mathrm{max}} \right] }
\end{equation}
\noindent where $N$ is the total number of vortices involved, and $\delta r$ is the bin size of the histogram. An empirical denominator is introduced to compensate for the RDF cut-off at large $r$ due to a finite image size. The four free parameters of the fit are $R_{1}$, $\sigma$, $R_{\mathrm{max}}$, and $W_{\mathrm{max}}$; their values are given in Table~\ref{tab:ionsize}. From equation (\ref{eq:zetadef}), we compute the correlation length of the vortex lattice $\zeta$. Here the larger $\zeta$ = 52 nm in Ba$_{0.6}$K$_{0.4}$Fe$_{2}$As$_{2}$\cite{shan2011observation} can be explained by the smaller ion size mismatch, whereas the small $\zeta$ = 22 nm in BaFe$_{1.8}$Co$_{0.2}$As$_{2}$\cite{yin2009scanning} and $\zeta$ = 20 nm in Sr$_{0.75}$K$_{0.25}$Fe$_{2}$As$_{2}$ are due to the large ion size mismatches there. Our hypothesis about the importance of dopant size mismatch to vortex pinning is further supported in the FeSe$_{x}$Te$_{1-x}$ system, where there is a large size mismatch between Se$^{2-}$ and Te$^{2-}$. Although a homogeneous superconducting gap and ordered vortex lattice are demonstrated in stoichiometric FeSe,\cite{song2011direct} both nanoscale chemical phase separation\cite{he2011nanoscale} and a disordered vortex arrangement\cite{hanaguri2010unconventional} are observed in FeSe$_{x}$Te$_{1-x}$.

Finally we comment on the quasiparticle bound states within the vortex cores,\cite{caroli1964bound} which often appear as a pronounced peak at or near $E_{\textrm{F}}$ in other Fe-SCs.\cite{shan2011observation, song2011direct, hanaguri2012scanning} No such states are observed in BaFe$_{1.8}$Co$_{0.2}$As$_{2}$\cite{shan2011observation} or Sr$_{0.75}$K$_{0.25}$Fe$_{2}$As$_{2}$ [Fig.\ 4(b)]. Using the residual resistivity $\rho_{0}=0.08$ $\mathrm{m}\Omega\cdot\mathrm{cm}$ and Hall coefficient $R_{\mathrm{H}}=1.16\times 10^{-9}$ $\mathrm{m}^{3}/\mathrm{C}$,\cite{chen2008transport, gen2008superconductivity} we obtain the electronic mean free path $\ell=\hbar(3\pi^{2})^{1/3}/e^{2}n^{2/3}\rho_{0}\sim5.2$ nm, two times bigger than $\xi\sim$ 2.3 nm. This suggests that our sample is macroscopically in the clean limit, where the vortex core bound states should have been observed. However, the vortices may be pinned in the relatively disordered regions, where the local mean free path is smaller.\cite{van2010quasiparticle} This indeed matches with the vortex pinning model hypothesized above. Moreover, we note that the visibility of vortex core bound states apparently depends on the surface structure within the same material Ba$_{0.6}$K$_{0.4}$Fe$_{2}$As$_{2}$.\cite{shan2011observation, wang2012close}

\section{\label{sec:Sum}Summary}
Our detailed STM/STS study of surface structure, superconducting gap, and vortex arrangement in Sr$_{0.75}$K$_{0.25}$Fe$_{2}$As$_{2}$ has addressed four important questions on Fe-SCs. First, images and spectroscopy of large patches of unreconstructed $1 \times 1$ surface provide the final unambiguous evidence that both $1\times 2$ and $\sqrt{2}\times\sqrt{2}$ reconstructions seen before represent a partial Sr layer. We observe no superconducting gap on the $1 \times 1$ As surfaces, in contrast to the ubiquitous gap on $1 \times 2$ surfaces, which reiterates the importance of attention to surface details when using STM/STS to study bulk superconductors. Second, our spatially resolved spectroscopy shows gap variation on a 3 nm length scale, larger than the 1.1 nm average distance between individual K dopants. This supports a K clustering model. Third, we have imaged a vitreous vortex phase with a short-range hexagonal order ($\zeta\sim$ 20 nm), suggesting strong pinning. Our hypothesis of the importance of dopant size mismatch suggests a way to optimize vortex pinning in Fe-SCs. Fourth, vortex core fitting gives a superconducting coherence length of $\xi\sim 2.3$ nm with no detectable anisotropy.

\begin{acknowledgments}
%\section{\label{sec:Sum}acknowledgments}
 We thank D.\ Inosov, M.\ C.\ Marchetti, and D.\ Nelson for helpful conversations. This work was supported by the Air Force Office of Scientific Research under grant FA9550-05-1-0371, and the U.S.\ National Science Foundation under grant DMR-0508812. C.\ L.\ S.\ was supported by the Golub Fellowship at Harvard University.
\end{acknowledgments}

%\bibliography{Sr122}
%

\end{document}